\documentclass[twoside,11pt]{article}

\usepackage[accepted]{melba}

\usepackage[utf8]{inputenc}
\usepackage{siunitx}
\usepackage[english]{babel}
\usepackage{color}
\usepackage[dvipsnames]{xcolor}
\usepackage{graphicx}
\usepackage{float}
\usepackage{subcaption}
\usepackage{ulem}
\usepackage{amsmath}
\usepackage{amssymb}
\usepackage{amsfonts}
\usepackage{hyperref}

\renewcommand{\subsubsection}{\textbf}
\usepackage{enumitem}

\usepackage[textwidth=2.0cm, textsize=tiny]{todonotes}

\newcommand{\Review}[1]{\textcolor{black}{#1}}

\usepackage{amsmath,amsfonts}

\melbaheading{2022:010}{https://www.melba-journal.org/papers/2022:010.html}{2021}{1-27}{01/2022}{03/2022}{Jallais, Rodrigues, Gramfort and Wassermann}{Information Processing in Medical Imaging (IPMI) 2021}{Aasa Feragen, Stefan Sommer, Julia Schnabel, Mads Nielsen}

\ShortHeadings{Inverting brain grey matter models with likelihood-free inference}{Jallais et al.}
\firstpageno{1}

\title{Inverting brain grey matter models with likelihood-free inference: a tool for trustable cytoarchitecture measurements}

\author{\name Ma\"{e}liss Jallais \email maeliss.jallais@inria.fr \\
	\addr Université Paris-Saclay, Inria, CEA, Palaiseau, France
	\AND
    \name Pedro L. C. Rodrigues \email pedro.rodrigues@inria.fr \\  
	\addr Univ. Grenoble Alpes, Inria, CNRS, Grenoble INP, LJK, Grenoble, France
	\AND
	\name Alexandre Gramfort \email alexandre.gramfort@inria.fr \\
	\addr Université Paris-Saclay, Inria, CEA, Palaiseau, France
	\AND
	\name Demian Wassermann \email demian.wassermann@inria.fr \\
	\addr Université Paris-Saclay, Inria, CEA, Palaiseau, France		
}

\begin{document}

\maketitle

\begin{abstract}
Effective characterisation of the brain grey matter cytoarchitecture with quantitative sensitivity to soma density and volume remains an unsolved challenge in diffusion MRI (dMRI). Solving the problem of relating the dMRI signal with cytoarchitectural characteristics calls for the definition of a mathematical model that describes brain tissue via a handful of physiologically-relevant parameters and an algorithm for inverting the model. To address this issue, we propose a new forward model, specifically a new system of equations, requiring a few relatively sparse $b$-shells.
We then apply modern tools from Bayesian analysis known as likelihood-free inference (LFI) to invert our proposed model. As opposed to other approaches from the literature, our algorithm yields not only an estimation of the parameter vector $\boldsymbol{\theta}$ that best describes a given observed data point $\boldsymbol{x}_0$, but also a full posterior distribution $p(\boldsymbol{\theta}|\boldsymbol{x}_0)$ over the parameter space. This enables a richer description of the model inversion, providing indicators such as credible intervals for the estimated parameters and a complete characterization of the parameter regions where the model may present indeterminacies. We approximate the posterior distribution using deep neural density estimators, known as normalizing flows, and fit them using a set of repeated simulations from the forward model. We validate our approach on simulations using \texttt{dmipy} and then apply the whole pipeline on two publicly available datasets.
\end{abstract}


\section{Introduction}

Obtaining quantitative measurements of brain grey matter microstructure with a dedicated soma representation is a growing field of interest in the diffusion MRI (dMRI) community \citep{PALOMBO_SANDI_2020,jelescu_NEXI_2021}. 
Unlike histology, dMRI permits to quantify brain tissue characteristics non-invasively and could, for example, help understanding dementia and cognitive deficits \citep{Douaud_cognitive_microstructure_2013}. However, current methods require demanding acquisitions with several $q$-shells (equivalently $b$-shells) and rely on non-linear models for which several parameter values may yield the same observation, also known as parameter indeterminacy~\citep{jelescu_degeneracy_2016,novikov_rotationally-invariant_2018}. This leads to numerically unstable solutions which are hard to interpret. Another major challenge is the fact that quantifying brain tissue microstructure directly from the dMRI signal is an inherently difficult task, because of the large dimensionality of the collected data and its stochastic nature~\citep{fick_mapl:_2016}.

Many works have tackled the non-linear inverse problem of inferring brain tissue parameters from dMRI signal. A popular solution is NODDI~\citep{zhang_noddi_2012}, which stabilises the solution by imposing constraints on model parameters that are not biologically plausible~\citep{novikov_rotationally-invariant_2018}. This has the effect of biasing the parameter estimation and the inverse problem remains largely degenerate \citep{jelescu_degeneracy_2016}. There has also been some interest in applying methods from the machine learning literature to solve the inverse problem. This is the case of the SANDI model \citep{PALOMBO_SANDI_2020}, in which the authors employ random forest regressors. Although the method provides rather acceptable accuracy in real-case scenarios, it can only output one set of tissue parameters for a given observed dMRI signal, masking, therefore, other biologically plausible solutions that could generate the same observed signal. \Review{Furthermore, the parameter estimates are obtained following a deterministic approach, so no description in terms of confidence interval are available. Also, the code for SANDI is not standard in the literature nor available in the internet yet.}

To overcome such limitations, we present three contributions. First, we use a three-compartment model for brain tissue \Review{composed of neurites, somas, and extra-cellular space}, and introduce a new parameter that jointly encodes soma radius and \Review{intracellular} diffusivity without imposing constraints on these values. This new parameter reduces indeterminacies in the model and has relevant physiological interpretations. Second, we present a method to fit the model through summary features of the dMRI signal based on a large and small $q$-value analysis using boundary approximations. These rotationally-invariant features relate directly to the tissue parameters and enable us to invert the model without manipulating the raw dMRI signals. Such summary statistics ensure a stable solution of the parameter estimations, as opposed to the indeterminate models used in~\citet{zhang_noddi_2012,PALOMBO_SANDI_2020}. Third, we employ modern tools from Bayesian analysis known as likelihood-free inference (LFI, ~\citep{Cranmer2020}) to solve our non-linear inverse problem under a probabilistic framework and determine the posterior distribution of the fitted parameters. Such approach offers a full description of the solution landscape and can point out degeneracies, as opposed to the usual deterministic least-squares based solution \citep{jelescu_degeneracy_2016,novikov_rotationally-invariant_2018}.




\section{Related Works}

Current brain tissue models are predominantly based on the two compartment Standard Model~(SM)~\citep{zhang_noddi_2012,novikov_quantifying_2018}. Recent evidence shows that the SM, mainly used in white matter, does not hold for grey matter microstructure analysis \citep{veraart_noninvasive_2020}. Several assumptions aim at explaining this issue such as increased permeability in neurite membranes~\citep{veraart_noninvasive_2020}, or curvy projections along with longer pulse duration~\citep{novikov_quantifying_2018}. We follow the hypothesis that the SM doesn't hold due to an abundance of cell bodies in grey matter \citep{PALOMBO_SANDI_2020}. Our proposed biophysical model is then based on three compartments~\citep{PALOMBO_SANDI_2020}: neurites, somas, and extra-cellular space (ECS). Despite its increased complexity, the main advantage of such model is the possibility to jointly estimate the characteristic features of each compartment.

Inferring parameters of the brain tissue model directly from dMRI signals has proven to be a very challenging task and has motivated the development of new approaches to reduce the dimensionality of the data to be processed. For instance,  \citet{novikov_rotationally-invariant_2018} proposes the LEMONADE system of equations, based on a Taylor expansion of the diffusion signal \Review{and a set of rotationally invariant moments}. In this work, we extend the LEMONADE equations to a setting with a three compartment model and further develop the method so to extract more features from the observed signal. We call the resulting vector of features defined by these quantities the `summary statistics' of the dMRI signal.

The usual way of applying a Bayesian approach to solve non-linear inverse problems~\citep{Stuart2010} is to define two quantities: a prior distribution encoding initial knowledge of the parameter values (e.g. intervals which are physiologically relevant) and the likelihood function of the forward model being studied. One can then either obtain an analytic expression of the posterior distribution via Bayes' formula or use a Markov-Chain Monte Carlo (MCMC) procedure to numerically sample the posterior distribution~\citep{Gelman2013}. However, the likelihood function of complex models such as the one that we consider here is often very hard to obtain and makes the Bayesian approach rather challenging to use. Likelihood-free inference (LFI) bypasses this bottleneck by recurring to several simulations of the forward model using different parameters and learning an approximation to the posterior distribution from these examples~\citep{Cranmer2020}.

The first contributions on LFI are also known as approximate Bayesian computation (ABC) and have been applied to invert models from ecology, population genetics, and epidemiology~\citep{Sisson2018}. Some of the limitations of these techniques include the large number of simulations required for the posterior estimations and the need of defining a distance function to compare the results of two simulations. Recently, there has been a growing interest in the machine learning community in improving the limitations of ABC methods through deep generative modeling, i.e. neural network architectures specially tailored to approximate probability density functions from a set of examples~\citep{Goodfellow2016}. Normalizing flows~\citep{Papamakarios2019} are a particular class of such neural networks that have demonstrated promising results for likelihood-free inference in different research fields~\citep{Cranmer2020, Gonalves2020, Greenberg2019}.


\section{Theory}

In this section, we present the theoretical background underlying our three main contributions. \textbf{Section~\ref{sec:3-compartment-model}} describes the three-compartment model for brain grey matter. We also introduce a new parameter that captures both soma radius and \Review{intracellular} diffusivity and avoids the usual indeterminacy in estimating them separately.  \textbf{Section~\ref{sec:summary-statistics}} presents the summary statistics used to reduce the dimensionality of the dMRI signal into a 7-dimensional feature vector that can be used to determine the physiological parameters of interest. \textbf{Section~\ref{sec:LFI}} describes the Bayesian approach used to solve our non-linear inverse problem and obtain a posterior distribution for the physiological parameters that best describe a given dMRI signal.

\subsection{Modeling the Brain grey Matter with a 3-compartment Model}
\label{sec:3-compartment-model}
To characterize cortical cytoarchitecture, we propose a method that relates the diffusion MRI signal to specific tissue parameters. To that aim, we first define a forward model based on a biophysical modeling of brain grey matter. 

Research in histology has demonstrated that grey matter is composed of neurons embedded in a fluid environment. Each neuron is composed of a soma, corresponding to the cellular body, surrounded by neurites connecting neurons together. Following this tissue biophysical composition, we model the grey matter tissue as three-compartmental \citep{PALOMBO_SANDI_2020}, moving away from the usual standard model (SM) designed for white matter. We further assume that our acquisition protocol is insensitive to exchanges between the compartments, i.e. molecules moving from one compartment to another have a negligible influence on the signal \citep{PALOMBO_SANDI_2020}. Many works include also a dot compartment into the SM with zero apparent diffusivity and no exchange. However, we have not considered such assumption, because its presence has been considered very unlikely in grey matter by previous works \citep{veraart_scaling_2019,Tax_dot_compartment_2020}. The observed diffusion signal is considered as a convex mixture of signals arising from somas, neurites, and extra-cellular space (ECS). Unlike white matter-centric methods~\citep{Jelescu2017}, we are not interested in the fiber orientation and only estimate orientation-independent parameters. This enables us to work on the direction-averaged dMRI signal, denoted $\bar{S}(q)$, known as the powder averaged signal. This consideration mainly matters for neurites, as their signal is not isotropic, as opposed to the proposed model for somas and ECS. Our direction-averaged grey matter signal model is then:
\begin{equation}
    \label{eq:S(q)}
    \frac{\bar{S}(q)}{S(0)} = {f_{n}}\bar{S}_{\mathrm{neurites}}({q}, {D_n}) + {f_{s}}\bar{S}_{\mathrm{somas}}({q}, {D_{s}}, {r_{s}}) + {f_{\mathrm{ECS}}} \bar{S}_{\mathrm{ECS}}({q}, D_{e})~.
\end{equation}
In this equation, $f_{n}$, $f_{s}$, and $f_{ECS}$ represent signal fractions for neurites, somas, and ECS respectively ($f_{n} + f_{s} +f_{\mathrm{ECS}} =1$). Note that the relative signal fractions do not correspond to the relative volume fractions of the tissue compartments as they are also modulated by different T2 values \citep{novikov_quantifying_2018}. $D_n$ corresponds to axial diffusivity inside neurites, while $D_{s}$ and $D_{e}$ correspond to somas and extra-cellular diffusivities. $r_{s}$ is the average soma radius within a voxel. This model is the same as the model SANDI proposed by \citet{PALOMBO_SANDI_2020}, with $f_n = f_{ic} f_{in}$, $f_s = f_{ic} f_{is}$ and $f_{\mathrm{ECS}} = f_{ec}$. We use $q$-values for more readability and harmonization throughout the paper, but a direct conversion to $b$-values is also possible, using $b = (2 \pi q)^2 \tau$ with $\tau = \Delta - \delta/3$.

We now review the model for each compartment, to make explicit the impact of each parameter on the diffusion MRI signal. \Review{Note that the use of such a model supposes each voxel contains a single type of neuron.}

\subsubsection{Neurite compartment.} Neurites, as in the SM, are modeled as 0-radius impermeable cylinders (``sticks"), with effective diffusion along the parallel axis, and a negligible radial intra-neurite diffusivity. In our acquisition setting, this model has been shown to hold~\citep{veraart_noninvasive_2020}. Its direction averaged signal is~\citep{veraart_noninvasive_2020}:
\begin{equation}
    \label{eq:power_law}
    \bar{S}_{\mathrm{neurites}}(q) \simeq \frac{1}{4 \sqrt{\pi \tau {D_n}}} \cdot q^{-1}~.
\end{equation}

\subsubsection{Soma compartment.} Somas are modeled as spheres, whose signal can be computed using the Gaussian phase distribution approximation \citep{balinov_nmr_1993}:
\begin{equation}
    -\log \bar S_{\mathrm{somas}}(q) = C({r_{s}}, {D_{s}}) \cdot q^2~.
\end{equation}
We exploit this relation here to extract a parameter $C_s = C(r_{s},D_{s}) [m^{2}]$ which, at fixed diffusivity $D_s$, is modulated by the radius of the soma $r_s$: 
\begin{align*}
    C({r_s}, {D_s}) = & \frac{2}{{D_s} \delta^2} \sum_{m=1}^{\infty} \frac{\alpha_m^{-4}}{\alpha_m^2 {r_s}^2 - 2} \\
    & \cdot \left( 2 \delta - \frac{2 + e^{-\alpha_m^2 {D_s} (\Delta - \delta)} - e^{-\alpha_m^2 {D_s} \delta} - e^{-\alpha_m^2 {D_s} \Delta} + e^{-\alpha_m^2 {D_s} (\Delta + \delta)}}{\alpha_m^2 {D_s}} \right)~,
    \label{eq:Cs}
\end{align*}
where $\alpha_m$ is the $m$th root of $(\alpha r_s)^{-1} J_{\frac{3}{2}}(\alpha r_s) = J_{\frac{5}{2}}(\alpha r_s)$, with $J_n(x)$ the Bessel functions of the first kind. In certain specific cases, $C_s$ has a simpler and more interpretable expression. For instance, when we consider a narrow pulse regime, with small $\tau$ or large $r_s$, $C_s$ loses its dependence on $r_s$ \citep{balinov_nmr_1993} and we obtain:
\begin{equation}
    C_s = D_s \tau~.
\end{equation}
In the Neuman (wide pulse) regime, i.e. when $D_s \Delta \gg r_s^2$ and $D_s \delta \ll 1$, $C_s$ becomes only dependent on the soma radius \citep{murday_cotts_1968}:
\begin{equation}
    C_s = \frac{1}{5} r_s^2~.
\end{equation}
These two approximations permits us to better interpret the parameter $C_s$.

\subsubsection{Extra-cellular space compartment.} \Review{We approximate the extra-cellular space with isotropic Gaussian diffusion}, i.e. a mono-exponential diffusion signal with a scalar diffusion constant $D_{e}$, which reflects the molecular viscosity of the fluid. This approximation assumes that the ECS is fully connected. The approximation is therefore:
\begin{equation}
    - \log(\bar{S}_{\mathrm{ECS}}(q)) = (2 \pi q)^2 \tau D_{e}~.
\end{equation}
Because of the geometry of the problem, we estimate $D_e$ as equal to one third of the diffusivity in the ventricles (considered as free diffusivity), given the same metabolic composition of the extracellular fluid and ventricles \citep{Vincent_sucrose_mouse_2021}.


\subsection{An Invertible 3-compartment Model: dMRI Summary Statistics}
\label{sec:summary-statistics}

The tissue model presented in Section \ref{sec:3-compartment-model} enables us to relate the dMRI signal with parameters representing grey matter tissue microstructure. However, solving the inverse problem directly from Eq.~\eqref{eq:S(q)} is a difficult task, leading to indeterminacies and poorly estimated parameters with large variability. Current methods addressing this issue have not studied its stability~\citep{PALOMBO_SANDI_2020} but simpler models with only two compartments have been shown to be indeterminate~\citep{novikov_quantifying_2018}.

To produce a method which addresses this indeterminacy, we introduce rotationally invariant summary statistics to describe the dMRI signal. The goal \Review{is} to reduce the dimensionality of the data at hand and representing all the relevant information for statistical inference with a few features. We then solve the inverse problem efficiently using Bayesian inference as described in Section~\ref{sec:LFI}. These dMRI-based summary statistics are extracted from our proposed model presented in Section~\ref{sec:3-compartment-model} via the following analysis of the dMRI signal on the boundaries of large and small $q$-value cases.

\subsubsection{Large $q$-value approximation: RTOP.}
We compute a $q$-bounded return-to-the-origin probability (RTOP), which measures the restrictions of the diffusing fluid molecule motion and gives us information about the structure of the media \citep{mitra_pulsed-field-gradient_1995}:
\begin{equation}
    \label{eq:RTOP_qmax}
    \mathrm{RTOP}(q) =  4\pi \int_{0}^{q}{\frac{\bar S(\eta)}{S(0)}} \eta^2 d\eta~.
\end{equation}
For $q$ large enough, the RTOP in our 3-compartment model in Eq.~\eqref{eq:S(q)} yields a soma and extra-cellular signal which converges towards a constant value in $q$, while the neurites' contribution becomes quadratic in $q$. In this case, RTOP becomes:
\begin{equation}
    \label{eq:RTOP_theoric}    
    \mathrm{RTOP}(q) = \underbrace{{f_s} \left( \frac{\pi}{{C_s}} \right)^{3/2} + \frac{{f_{\mathrm{ECS}}}}{ 8 (\pi \tau D_{e})^{3/2}}}_{a_\mathrm{fit}} + \underbrace{\frac{{f_{n}}}{2} \cdot \sqrt{\frac{\pi}{\tau {D_n}}}}_{b_\mathrm{fit}} \cdot q^2 + \gamma q^3~,
\end{equation}
in which the last term of the equation, $\gamma q^3$, is a nuisance parameter that describes a constant noise in the direction averaged signal. By accurately estimating the second derivative of $\mathrm{RTOP}(q)$ at $q$ large enough, we can solve the coefficients of interest of the polynomial in Eq.~\eqref{eq:RTOP_theoric}: $a_{\mathrm{fit}}$ and $b_{\mathrm{fit}}$.  We do this efficiently by casting it as an overdetermined ordinary least squares problem which has a unique solution. 

\subsubsection{Small $q$-value approximation: Spiked LEMONADE\footnote{\Review{Please note that the `spiked' adjective used here is simply an allusion to our method being a variation of the LEMONADE method.}}.}
\label{sec:Lemonade}
We propose a second approximation, based on a moment decomposition for small $q$-values~\citep{novikov_rotationally-invariant_2018}:
\begin{equation}
    \frac{S_{\hat{\textbf{g}}}(q)}{S(0)} = 1 - b(q) M^{(2)}_{i_1 i_2} g_1 g_2 + \frac{b(q)^2}{2!} M^{(4)}_{i_1\ldots i_4} g_1\ldots g_4 - \ldots,\quad b(q)=(2\pi q)^2\tau~,
    \label{eq:moments}
\end{equation}
where $i_k$ are the directional basis of the tensors $M$,  $g_k = i_k \cdot \hat{\textbf{g}}\in \mathbb{R}^3$, and $\hat{\textbf{g}}$ the unit direction of the dMRI acquisition. From the moment tensors of this decomposition, LEMONADE~\citep{novikov_rotationally-invariant_2018} extracts rotational invariant scalar indices $M^{(i),j},\,i, j\in \{0, 2, 4,\ldots\}.$ These quantify white matter microstructure by plugging the 2-compartment SM into Eq.~\eqref{eq:moments}~\citep[see][app. C]{novikov_rotationally-invariant_2018}.

In this work, we extended LEMONADE to our 3-compartment model presented in Section~\ref{sec:3-compartment-model} by plugging Eq.~\eqref{eq:S(q)} into Eq.~\eqref{eq:moments} and performing tedious arithmetic. This results in the following equation system, which now includes the soma parameter $C_s$, relating the dMRI signal to grey matter microstructure:
\begin{equation}
    \label{eq:moment-ms}
    \begin{cases}
        M^{(2),0} = {f_n} {D_n} + 3 {f_s} \frac{{C_s}}{(2 \pi)^2 \tau} + 3 {f_{\mathrm{ECS}}} D_e \\
        M^{(2),2} = {f_n} {D_n}{p_2} \\
        M^{(4),0} = {f_n} {D_n}^2 + 5 {f_s} \left(\frac{{C_s}}{(2 \pi)^2 \tau}\right)^2 + 5 {f_{\mathrm{ECS}}} D_e^2 \\
        M^{(4),2} = {f_n} {D_n}^2{p_2}
    \end{cases}
\end{equation}
where $p_2$ is a scalar measure of neurite orientation dispersion~\citep{novikov_rotationally-invariant_2018}.

Note that only the shells with $b(q)\leq$ \SI{2.5}{\milli\second\per\square\micro\meter} are used, to get an unbiased estimation of the rotational invariant moments $M^{(2),0}$, ${M^{(2),2}}$, ${M^{(4),0}}$ and ${M^{(4),2}}$.

\subsubsection{Complete system.}
Combining equations \eqref{eq:RTOP_theoric} and \eqref{eq:moment-ms} and adding the constraint that the fractions for the three compartments sum to one, we obtain a non-linear system of 7 equations and 7 unknowns. Following \citet{menon2020microstructural}, we assume $D_e$ nearly-constant per subject acquisition and estimated it as one third of the mean diffusivity in the subject's ventricles~\citep{Vincent_sucrose_mouse_2021}. This assumption allows us to drop an unknown from the system, use $D_e$ as a reference diffusivity and \Review{turn our system of equations unitless with} ${D_n^{u}} = \frac{{D_n}}{D_e}$ and ${C_s^{u}} = \frac{{C_s}}{(2\pi)^2 \tau D_e}$, which gives:
\begin{center}
\begin{tabular}{|c|c|}
\hline \textbf{Spiked LEMONADE} & \textbf{RTOP} \\
Small $q$-values & Large $q$-values \\
\hline $\begin{cases}
    \frac{M^{(2),0}}{D_e} = {f_n} {D_n^{u}} + 3 {f_s} {C_s^{u}} + 3 {f_{\mathrm{ECS}}} \\
    \frac{M^{(2),2}}{D_e} = {f_n} {D_n^{u}} \cdot {p_2} \\
    \frac{M^{(4),0}}{D_e^2} = {f_n} {D_n^{u}}^2 + 5 {f_s} {C_s^{u}}^2 + 5 {f_{\mathrm{ECS}}} \\
    \frac{M^{(4),2}}{D_e^2} = {f_n} {D_n^{u}}^2 \cdot {p_2}\\
\end{cases}$
& $\begin{cases}
    a_{\mathrm{fit}} \left(\tau D_e\right)^{3/2}=& \frac{{f_s}}{8(\pi {C_s^u})^{3/2}} \\
    &+ \frac{{f_{\mathrm{ECS}}}}{ 8 \pi^{3/2} } \\
    b_{\mathrm{fit}} \left(\tau D_e\right)^{1/2} =& \frac{{f_{n}}}{2} \sqrt{\frac{\pi}{{D_n^u}}}
\end{cases}$ \\
\hline
\multicolumn{2}{|c|}{${f_n} + {f_s} + {f_{\mathrm{ECS}}} = 1$} \\
\hline
\end{tabular}
\end{center}



\subsection{Inverting the model with Bayesian inference}
\label{sec:LFI}

Our main goal is to determine the values of the parameter vector $$\boldsymbol{\theta} = (D_n, C_s, p_2, f_s, f_n, f_\mathrm{ECS}) \in \mathbb{R}^{6}$$ that best explain a given observed dMRI signal. Because of the high-dimensionality of this kind of signal and the difficulties in obtaining stable estimates of $\boldsymbol{\theta}$ directly from it, we recur to the set of summary features defined in Section~\ref{sec:summary-statistics},
$$\boldsymbol{x} = \left(\tfrac{M^{(2),0}}{D_e}, \tfrac{M^{(2),2}}{D_e}, \tfrac{M^{(4),0}}{D_e^2},  \tfrac{M^{(4),2}}{D_e^2}, a_{\mathrm{fit}} (\tau D_e)^{3/2}, b_{\mathrm{fit}} \sqrt{\tau D_e}, 1\right) \in \mathbb{R}^{7}~,$$ and make the assumption that it carries all the information necessary for determining the $\boldsymbol{\theta}_0$ having generated a given dMRI signal $\boldsymbol{S}_0$. We denote the relation between these quantities as
\begin{equation}
\label{eq:input-output-relation}
\boldsymbol{x} = \mathcal{M}(\boldsymbol{\theta}) + \boldsymbol{n}~,
\end{equation}
where $\mathcal{M}:\mathbb{R}^{6} \to \mathbb{R}^{7}$ is a multivariate function that implements the system of equations described in Section~\ref{sec:summary-statistics} and $\boldsymbol{n}$ is an additive noise capturing the imperfections of our modelling procedure, the limitations of the summary statistics, and the measurement noise.

{\subsubsection{The Bayesian formalism.} 
We interpret the inverse problem of inferring the parameters that best describe a given observed summary feature vector $\boldsymbol{x}_0$ as that of determining the posterior distribution of $\boldsymbol{\theta}$ given an observation $\boldsymbol{x}_0$. By first choosing a prior distribution $p(\boldsymbol{\theta})$ describing our initial knowledge of the parameter values, we may use Bayes' theorem to write
\begin{equation}
\label{eq:bayes-theorem}
p(\boldsymbol{\theta}|\boldsymbol{x}_0) = \dfrac{p(\boldsymbol{x}_0|\boldsymbol{\theta})p(\boldsymbol{\theta})}{p(\boldsymbol{x}_0)} \enspace , 
\end{equation}
where $p(\boldsymbol{x}_0|\boldsymbol{\theta})$ is the likelihood of the observed data point and $p(\boldsymbol{x}_0)$ is a normalizing constant, commonly referred to as the evidence of the data. Note that such a probabilistic approach returns not only which $\boldsymbol{\theta}$ best fits the observed data (i.e. the parameter that maximizes the posterior distribution), but the full posterior distribution $p(\boldsymbol{\theta}|\boldsymbol{x}_0)$. The latter can be possibly multimodal or flat, which would indicate the difficulty of summarizing it with a unique maximum.

\subsubsection{Bypassing the likelihood function.}
Despite its apparent simplicity, it is usually difficult to use Eq.~\eqref{eq:bayes-theorem} to determine the posterior distribution, since the likelihood function for data points generated by complex non-linear models is often hard to write. To avoid such difficulty, we directly approximate the posterior distribution using a conditional density estimator, i.e. a family of conditional p.d.f. approximators $q_{\phi}(\boldsymbol{\theta}|\boldsymbol{x})$ parametrized by $\phi$ and that takes $\boldsymbol{\theta}$ (the parameter) and $\boldsymbol{x}$ (the observation) as input arguments. Our posterior approximation is then obtained by minimizing its average Kullback-Leibler divergence with respect to the conditional density estimator for different choices of $\boldsymbol{x}$, as per~\citep{Papamakarios2016}
\begin{equation}
\label{eq:loss-function-minimizer}
\begin{array}{llll}
\underset{\boldsymbol{\phi}}{\mbox{min.}} & \mathcal{L}(\phi) & \text{with} & \mathcal{L}(\phi) = \mathbb{E}_{\boldsymbol{x} \sim p(\boldsymbol{x})}\left[D_{\text{KL}}(p(\boldsymbol{\theta}|\boldsymbol{x}) \| q_{\phi}(\boldsymbol{\theta}|\boldsymbol{x}))\right]~,
\end{array}
\end{equation}
which can be rewritten as
\begin{equation}
\begin{array}{rcl}
\mathcal{L}(\phi) &=& {\displaystyle\int} D_{\text{KL}}(p(\boldsymbol{\theta}|\boldsymbol{x}) \| q_{\phi}(\boldsymbol{\theta}|\boldsymbol{x}))p(\boldsymbol{x})\mathrm{d}\boldsymbol{x}~, \\[0.90em]
&=& -{\displaystyle\iint}\log\big(q_{\phi}(\boldsymbol{\theta}|\boldsymbol{x})\big)p(\boldsymbol{\theta}|\boldsymbol{x})p(\boldsymbol{x})\mathrm{d}\boldsymbol{\theta}\mathrm{d}\boldsymbol{x} + C~, \\[0.90em]
&=& -{\displaystyle\iint}\log\big(q_{\phi}(\boldsymbol{\theta}|\boldsymbol{x})\big)p(\boldsymbol{x}, \boldsymbol{\theta})\mathrm{d}\boldsymbol{\theta}\mathrm{d}\boldsymbol{x} + C~, \\[0.90em]
&=& -\mathbb{E}_{(\boldsymbol{x}, \boldsymbol{\theta}) \sim p(\boldsymbol{x}, \boldsymbol{\theta})}\left[\log\big(q_{\phi}(\boldsymbol{\theta}|\boldsymbol{x})\big)\right] + C~,
\end{array}
\end{equation}
where $C$ is a constant that does not depend on $\phi$. Note, however, that in practice we actually consider a $N$-sample Monte-Carlo approximation of the loss function given by
\begin{equation}
\label{eq:loss-function}
\mathcal{L}(\phi) \approx \mathcal{L}^N(\phi) = -\frac{1}{N}\sum_{i = 1}^{N}\log\Big(q_{\phi}(\boldsymbol{\theta}_i|\boldsymbol{x}_i)\Big)~,
\end{equation}
where the $N$ data points $(\boldsymbol{\theta}_i, \boldsymbol{x}_i)$ are sampled from the joint distribution with $\boldsymbol{\theta}_i \sim p(\boldsymbol{\theta})$ and $\boldsymbol{x}_i \sim p(\boldsymbol{x}|\boldsymbol{\theta}_i)$. We can then use stochastic gradient descent to obtain a set of parameters $\phi$ which minimizes $\mathcal{L}^N$.
}
If the class of conditional density estimators is sufficiently expressive, it is possible to show that the minimizer of Eq.~\eqref{eq:loss-function} converges to $p(\boldsymbol{\theta}|\boldsymbol{y})$ when $N \to \infty$~\citep{Greenberg2019}. Note, also, that the parametrization $\phi$ that we obtain by the end of the optimization procedure yields a posterior which is amortized for different choices of $\boldsymbol{x}$. Thus, for a specific observation $\boldsymbol{x}_0$ we may simply write $ q_\phi(\boldsymbol{\theta}|\boldsymbol{x}_0)$ to get an approximation of $p(\boldsymbol{\theta}|\boldsymbol{x}_0)$.

\subsubsection{Neural density estimators.} In this work, our conditional p.d.f. approximators belong to a class of neural networks called normalizing flows~\citep{Papamakarios2019}. These flows are invertible functions capable of transforming vectors generated by a simple base distribution (e.g. the standard multivariate Gaussian distribution) into an approximation of the true posterior distribution. An important advantage of normalizing flow versus other p.d.f. approximators such as generative adversarial network (GAN~\citep{Goodfellow2014}) and variational auto-encoders (VAE~\citep{Kingma2014}) is that it provides both the likelihood of any sample point of interest and it is also straightforward to sample new data points from it. Furthermore, certain classes of normalizing flows can be shown to be universal approximators of probability density functions. We refer the reader to \citet{Papamakarios2019} for more information on the different types of normalizing flows.

\section{Materials and methods}
This section presents the technical details on how we have implemented our theoretical contributions and describes the simulated and real datasets used in the numerical illustrations.

\subsection{The likelihood-free inference setup}
\label{sec:lfi-setup}
Using a likelihood-free inference approach for inverting the 3-compartment model relating tissue parameters $\boldsymbol{\theta}$ and dMRI summary statistics $\boldsymbol{x}$ relies on four important aspects:
\begin{enumerate}[itemsep=0.3em]
    \item[(1)] \textbf{The forward model.} As explained in Section~\ref{sec:LFI}, we obtain an approximation of the amortized posterior distribution using a dataset containing several paired examples of a parameter $\boldsymbol{\theta}_i$ and its corresponding summary statistics $\boldsymbol{x}_i$. In what follows, we adopt the usual assumptions from the inverse problems literature and consider the additive noise $\boldsymbol{n}$ from Eq.~\eqref{eq:input-output-relation} small enough to be ignored in the data generation, so that we have $\boldsymbol{x}_i \approx \mathcal{M}(\boldsymbol{\theta}_i)$.
    \item[(2)]\textbf{Prior distribution.} The simplest way of defining a prior distribution $p(\boldsymbol{\theta})$ in our setting is to use an uniform distribution with limits within physiologically relevant intervals for each parameter. From Section~\ref{sec:3-compartment-model}, we know that the fractions $f_s$, $f_n$, and $f_{\text{ECS}}$ have values between 0 and 1 and all sum up to one. To encode this information in $p(\boldsymbol{\theta})$, we define new parameters $k_1$ and $k_2$ and relate them with the fractions by
\begin{equation}
\label{eq:fractions-k1k2}
    f_n = k_2\sqrt{k_1}, \quad f_s = \sqrt{k_1}(1-k_2), \quad f_{\text{ECS}} = 1 - \sqrt{k_1}~.
\end{equation}
In this way, whenever we want to generate a prior sample for $(f_n, f_s, f_{\text{ECS}})$ we first generate a sample of $k_1, k_2 \sim \mathcal{U}([0, 1])$ and then transform them according to \eqref{eq:fractions-k1k2} to get a set of fractions which is uniformly sampled in the region $\{f_n , f_s, f_{\text{ECS}} \in [0, 1]~: f_n + f_s + f_{\text{ECS}} = 1\}$. We follow the usual assumption that the diffusivity of the compartments are inferior or equal to the self-diffusion coefficient of free water, which is \SI{3}{\square\micro\meter\per\milli\second}~\citep{Li2016}. We fix, therefore, the interval for neurite diffusivity ($D_n$) as between $10^{-5}$ and $3$ \si{\square\micro\meter\per\milli\second}. The newly introduced parameter $C_s$ is parametrized by soma radius and diffusivity. To account for a soma radius comprised between 2 and \SI{30}{\micro\meter} \citep{palombo_neuron_size_2021} and a diffusivity range as previously defined, we used the $C_s$ interval [50, 2500] \si{\square\micro\meter}. Parameter $p_2$, which measures the dispersion of neurites orientation, is comprised in the interval [0,1], with 1 indicating an anisotropic orientation distribution function.
\item[(3)]\textbf{Posterior approximator.} We use an autoregressive architecture for normalizing flows implemented via the masked autoencoder for distribution estimation (MADE)~\citep{Germain2015}. We follow the same setup from~\citet{Greenberg2019} for LFI problems, stacking five MADEs, each with two hidden layers of 50 units, and a standard normal base distribution. This choice provides a sufficiently flexible function capable of approximating complex posterior distributions.
\item[(4)]\textbf{Training procedure.} The parametrization of our normalizing flow is obtained by minimizing the loss function~\eqref{eq:loss-function} using the ADAM optimizer~\citep{Kingma2017} with default parameters, a learning rate of $5.10^{-4}$ and a batch size of 100. Except for a few validation experiments, we have used $N = 10^5$ simulated data points to approximate the posterior distribution.
\end{enumerate}

\subsection{Simulated dMRI data}

We first validate our proposed method using simulated dMRI data. For this, we fix a parameter vector $\boldsymbol{\theta}_0$ based on a plausible biophysical configuration~\citep{palombo_neuron_size_2021} and generate a simulated observation $\boldsymbol{x}_0$ associated to it. Our goal, then, is to check whether the posterior distribution $p(\boldsymbol{\theta}|\boldsymbol{x}_0)$ concentrates around the ground truth parameter, i.e. if it is peaked around the true values of the parameters in $\boldsymbol{\theta}_0$. If this is the case, we can assert that the LFI procedure is capable of inverting our non-linear model successfully.

The simplest way of generating an observation from the ground truth parameter would be to use the forward model defined in Section~\ref{sec:lfi-setup}, \Review{which yields very good results, since the posterior approximation is trained on data points generated in the same way. We have also considered a more challenging situation, in which the dMRI signals are simulated following a setup that is closer to what we would expect from real experimental experiments.} This is based on two steps. Firstly, we use the \texttt{dmipy} simulator \citep{fick_dmipy_2019} to simulate the three-compartment model described in Section~\ref{sec:3-compartment-model} and obtain a dMRI signal $\boldsymbol{S}_0$. Then, we calculate {the summary statistics of this signal as defined in Section~\ref{sec:summary-statistics}} to reduce the dimensionality of the observation and obtain a feature vector $\boldsymbol{x}_0$. 

We have carried out our simulations on \texttt{dmipy} considering three different kinds of acquisition setup. They all have $b$-shells with 128 uniformly distributed directions, but they differ in their $b$-values and acquisition times: 
\let\labelitemi\labelitemii
\begin{itemize}
    \item Setup \texttt{Ideal} corresponds to a rather ``confortable'' case with 10 $b$-values between 0 and \SI{10}{\milli\second\per\square\micro\meter}. We use $\delta/\Delta$ = \SI{12.9/21.8}{\milli\second} as in the HCP MGH database.
    \item Setup \texttt{HCP MGH} reproduces the setup from the HCP MGH dataset, with 5 $b$-values: 0, 1, 3, 5, and \SI{10}{\milli\second\per\square\micro\meter} and $\delta/\Delta$ = \SI{12.9/21.8}{\milli\second}. {Since the Spiked LEMONADE approximation (\ref{sec:Lemonade}) requires at least three $b$-values inferior to \SI{2.5}{\milli\second\per\square\micro\meter}, we extrapolated an extra $b$-shell at \SI{0.1}{\milli\second\per\square\micro\meter} \Review{using \texttt{MAPL}~\citep{fick_mapl:_2016}, a method for modeling multi-shell q-space signals.}}
    \item Setup \texttt{HCP 1200} reproduces the setup from the HCP 1200 dataset, with only 4 $b$-values: 0, 1, 2, and \SI{3}{\milli\second\per\square\micro\meter} and $\delta/\Delta$ = \SI{10.6/43.1}{\milli\second} An extra b-shell at \SI{2.8}{\milli\second\per\square\micro\meter} has been interpolated to be used in the RTOP approximation. 
\end{itemize}
{Note that in the simulations with all setups we have used the three $b$-shells with the lowest $b$-values for the small $q$-value approximation (Spiked LEMONADE), and the three largest $q$-values for the RTOP approximation.}

\subsection{HCP MGH dataset}
After validating our proposal on different simulated settings, we carried out our analysis on two publicly available databases. \Review{Our goal was to estimate the tissue parameters for each voxel in the dMRI acquisitions corresponding to the grey matter and determine how these parameters vary. We segmented the brain grey matter using \texttt{FreeSurfer} before applying our pipeline to the selected voxels.} Because of the probabilistic framework that we use, these estimates are accompanied of credible intervals that can be used to inform our degree of confidence of these estimates.

Our first analysis was on the HCP MGH Adult Diffusion database \citep{SETSOMPOP2013220}. This database is composed of 35 subjects with $\delta/\Delta=12.9/\SI{21.8}{\milli\second}$ and $b$ = 1, 3, 5, \SI{10}{\milli\second\per\square\micro\meter}. We used the 3, 5 and \SI{10}{\milli\second\per\square\micro\meter} $b$-values for the RTOP approximation (i.e. the large $q$-value analysis), and \Review{0, 0.1 and \SI{1}{\milli\second\per\square\micro\meter} for the Spiked LEMONADE approximation. We used \texttt{MAPL} with the 0, 1 and \SI{3}{\milli\second\per\square\micro\meter} $b$-values to reduce noise and interpolate a $b$-value of \SI{0.1}{\milli\second\per\square\micro\meter} to improve the estimations}. $D_e$ was estimated as $1/3$ of the mean diffusivity in the ventricles. 

{The spatial distribution of the estimated parameters were mapped to the MNI template, averaged over all subjects, and then projected onto an inflated cortical surface using \texttt{FreeSurfer}. We have then evaluated whether such distributions seemed physiologically plausible by using the Brodmann atlas~\citep{brodmann1909vergleichende,Zilles2018}, which is a parcellation of the brain based on cytoarchitecture features. In addition, we compared qualitatively the results of soma estimations with Nissl-stained histological images of cytoarchitecture~\citep{allmanEconomoNeuronsFrontoinsular2010,amuntsBrocaRegionRevisited,geyerAreas3a3b1999}.}

\subsection{HCP 1200}
We proceeded with our analysis on real data using a more challenging database, in which the dMRI signals were acquired with only a few \Review{small} $b$-values. Our goal was to demonstrate that the credible regions obtained via the posterior approximation can be used to inform which parameters remain possible to estimate even in very challenging situations. Note that this unlocks the door to the analysis of any dMRI database, since the estimates always come with a ``quality certificate''.

We applied our pipeline to a subset of the HCP 1200 database. We randomly picked 30 subjects, to have an identical number of subjects to that in our analysis of the HCP MGH database. The data were acquired for $b$-values equal to 1, 2 and  \SI{3}{\milli\second\per\square\micro\meter}, with $\delta/\Delta=10.6/\SI{43.1}{\milli\second}$. Using \texttt{MAPL}, we interpolated a b-shell at \SI{2.8}{\milli\second\per\square\micro\meter} to improve the computation of the summary statistics. We used all the three lowest $b$-values for the Spiked LEMONADE approximation, and $b$ = 2, 2.8 and \SI{3}{\milli\second\per\square\micro\meter} for the high $b$-value approximation based on RTOP. {Similarly to the HCP MGH dataset, we have averaged the parameter estimations in a common space (MNI) and then projected the resulting parametric maps onto an inflated cortical surface.}

\Review{\subsection{Software}
All our experiments have been implemented with Python~\citep{Python36} using several scientific packages: dMRI signals were simulated with the package \texttt{dmipy}~\citep{fick_dmipy_2019} {and processed using  \texttt{dipy}~\citep{dipy2014}} or custom implementations based on \texttt{numpy}~\citep{harris2020array}. We used the \texttt{sbi}~\citep{tejero-cantero2020sbi} and \texttt{nflows}~\citep{Durkan2020nflows} packages for carrying out the LFI procedures and combined them with data structures and functions from \texttt{pyTorch}~\citep{Paszke2019pyTorch}. The figures of results on real experimental data were generated with \texttt{mayavi}~\citep{ramachandran2011mayavi}.}

\section{Results}
In this section, we describe our results obtained on simulated data and two real datasets. 

\subsection{Simulated data}
\subsubsection{Validating the LFI pipeline.} In this first round of experiments, our aim was to check whether the LFI pipeline worked correctly on a setting where we knew the true values of the parameter $\boldsymbol{\theta}_0$ (ground truth) generating the observed data $\boldsymbol{x}_0$. Furthermore, we wanted to confirm whether the use of summary statistics for the dMRI signal actually conveyed any improvements in the parameter estimation. We have considered, therefore, three different cases:
\begin{itemize}[itemsep=0.3em]
\let\labelitemi\labelitemii
    \item \textbf{Case 1.} Generate $\boldsymbol{x}_0$ directly from $\boldsymbol{\theta}_0$ using the forward model defined in Section~\ref{sec:lfi-setup}. This is a rather favorable case for our posterior approximation, since it is applied on a data point generated in the same way as the dataset in which it was trained.
    \item \textbf{Case 2.} Generate a dMRI signal $\boldsymbol{S}_0$ from $\boldsymbol{\theta_0}$ using \texttt{dmipy} and then obtain $\boldsymbol{x}_0$ by calculating the summary statistics presented in Section~\ref{sec:summary-statistics}. We use the same posterior approximator from \textbf{Case 1}, meaning that the data point in inference time is generated differently from those for the training procedure. This is the actual realistic case that interests us the most.  
    \item \textbf{Case 3.} Generate a dMRI signal $\boldsymbol{S}_0$ from $\boldsymbol{\theta_0}$ using \texttt{dmipy} and do not use any summary statistics for the model inversion, i.e. consider $\boldsymbol{x}_0 = \boldsymbol{S}_0$. Note that the posterior approximator has to be trained on a dataset with observations consisting of dMRI signals, so it is different from the approximators in \textbf{Case 1} and \textbf{Case 2}. Depending on how the LFI pipeline behaves on this case, the use of summary statistics can be justified or not.     
\end{itemize}
All simulations were carried out with setup \texttt{Ideal}, which corresponds to an ideal dMRI acquisition scheme, and the posterior approximators were trained with $N = 10^5$ simulated data points. While we have validated the LFI pipeline on multiple choices of physiologically relevant ground truth parameters $\boldsymbol{\theta}_0$, \mbox{Figure \ref{fig:sim_base_case_setupA}} portrays the results only for
$$\boldsymbol{\theta}_0 = (D_n, C_s, p_2, f_s, f_n, f_\mathrm{ECS}) = (2.5~\si{\square\micro\meter\per\milli\second}, 617~\si{\square\micro\meter}, 0.50, 0.15, 0.45, 0.40)~.$$ This choice of parameters represent a sample tissue containing pyramidal neurons of radius \SI{12}{\micro\meter}~\citep{palombo_neuron_size_2021} and diffusivity \SI{3}{\square\micro\meter\per\milli\second}. The three compartment proportions were chosen in accordance to reported values from histology of human grey matter tissue~\citep{Shapson-Coe_GM_tissue}. Results are presented in Figure \ref{fig:sim_base_case_setupA}. We see that in \textbf{Case 1} the marginals of the posterior distribution are well concentrated around the values of the ground truth parameter $\boldsymbol{\theta}_0$. This confirms that the posterior approximator successfully inverts the non-linear model using the examples in the training set. We also see that the parameter estimation in \textbf{Case 2} captures very well most of the true values of the parameters, indicating both that our posterior approximator is robust to observed data generated differently from its training set {and that our summary statistics are descriptive enough to synthesize each tissue configuration}. Finally, the poor results in the estimation for \textbf{Case 3} reflect the largely indeterminate system of equations that results from trying to infer the tissue parameters directly from the dMRI signals. This behavior was expected, as similar degeneracies were shown for a simpler model in~\citet{novikov_quantifying_2018}.

\subsubsection{Influence of the number of $b$-values.} 
\begin{figure}
    \centering
    \includegraphics[width=0.65\textheight]{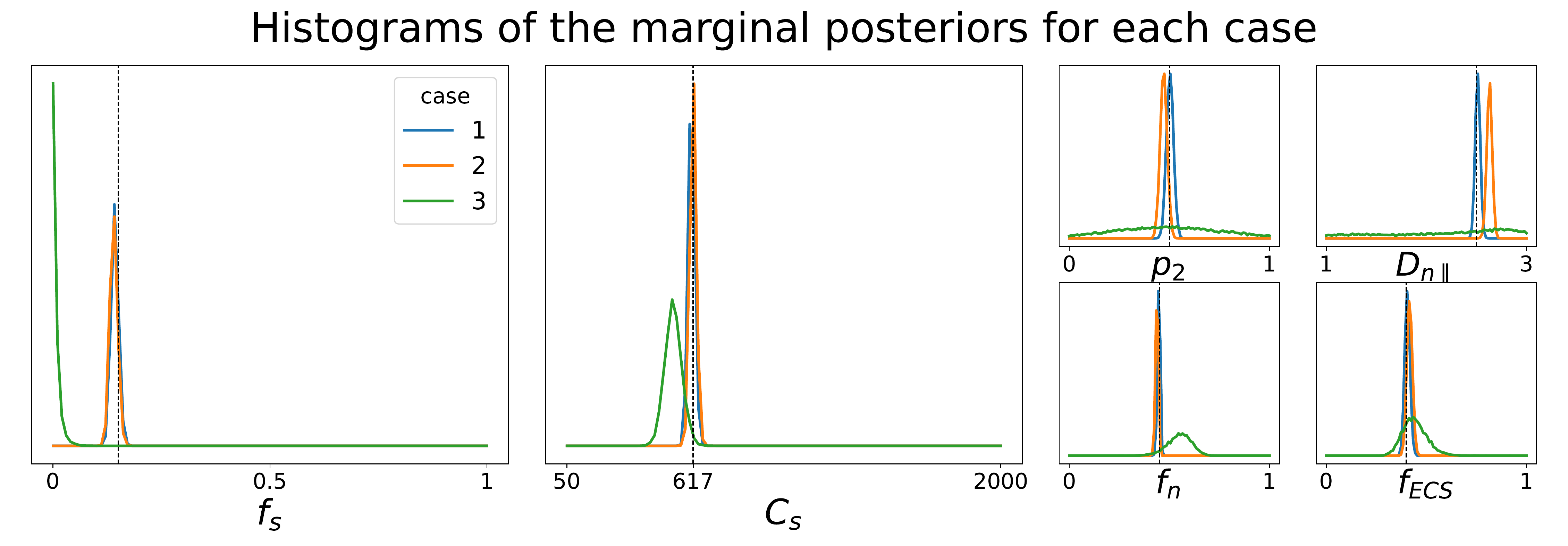}
    \caption{Histograms of $10^4$ samples of the approximate posterior distribution in three different base cases (see text for details). Vertical black dashed lines represent ground truth values of $\boldsymbol{\theta}_0$ which generated the observed dMRI signals. We observe that the marginals tend to concentrate around the ground truth parameters when the observed summary statistics are obtained directly from the parameters (\textbf{Case 1}) and have a small bias when the signals are generated using \texttt{dmipy} (\textbf{Case 2}). The figure also shows that inverting the model directly from the dMRI signals leads to rather poor results (\textbf{Case 3}).}
    \label{fig:sim_base_case_setupA}
\end{figure}

\begin{figure}
    \centering
    \includegraphics[width=0.65\textheight]{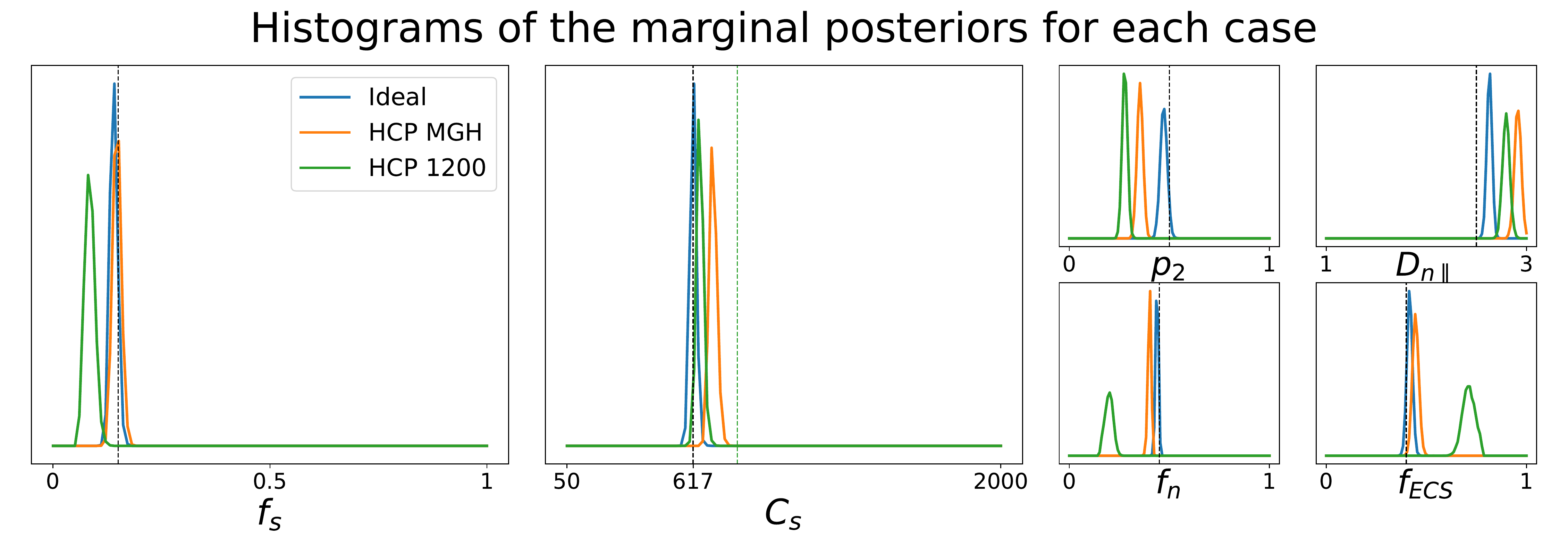}
    \caption{Histograms of $10^4$ samples of the approximate posterior distribution with observed dMRI signals generated under three acquisition setups (see text for details). Vertical black dashed lines represent ground truth values of $\boldsymbol{\theta}_0$ which generated the observed signals (the green dashed line indicates the $C_s$ value expected for the \texttt{HCP 1200} setup). We see that while the \texttt{Ideal} case delivers very good estimates, the results for the two other setups are only reliable for a subset of the tissue parameters.}
    \label{fig:sim_base_case}
\end{figure}

We have also investigated how different choices of $b$-values in the acquisition scheme affect the quality of the parameter estimation using the posterior approximation. Note that these choices have no influence on how the posterior approximator is obtained, since it is trained on data generated directly from the relations between tissue parameters and the diffusion summary statistics, in which the $b$-values do not interact. In fact, the different acquisition setups that we consider have only an impact over the observed data point generated via \texttt{dmipy}. Figure \ref{fig:sim_base_case} portrays the marginal posterior distributions for each tissue parameter in setups \texttt{Ideal}, \texttt{HCP MGH}, and \texttt{HCP 1200}. We see that estimations in the \texttt{Ideal} setup (equivalent to Case 2 in Figure \ref{fig:sim_base_case_setupA}) are very much concentrated around the true values of the parameters. \Review{For \texttt{HCP MGH} and \texttt{HCP 1200} the estimations of the soma-related parameters are rather good, but a small bias is present for the other parameters in \texttt{HCP MGH} and even more for \texttt{HCP 1200}.} Note that the \texttt{Ideal} and \texttt{HCP MGH} ground truth value of $C_s$ (vertical black dashed line) is different from the one of the \texttt{HCP 1200} setup (vertical green dashed line), because of their different diffusion times. These simulations allow us to have a fair confidence in the estimations on real data, presented in Sections \ref{sec:HCP_results} and \ref{sec:HCP_1200_results}.

\subsubsection{Our new parameter avoids model indeterminacy.}
In Section~\ref{sec:3-compartment-model}, we introduce the parameter $C_s$, which serves as a proxy of the soma radius and provides key information on the soma compartment. In this experiment, we illustrate the results of our model inversion if we had not defined parameter $C_s$. 

\begin{figure}
    \centering
    \includegraphics[width=0.65\textheight]{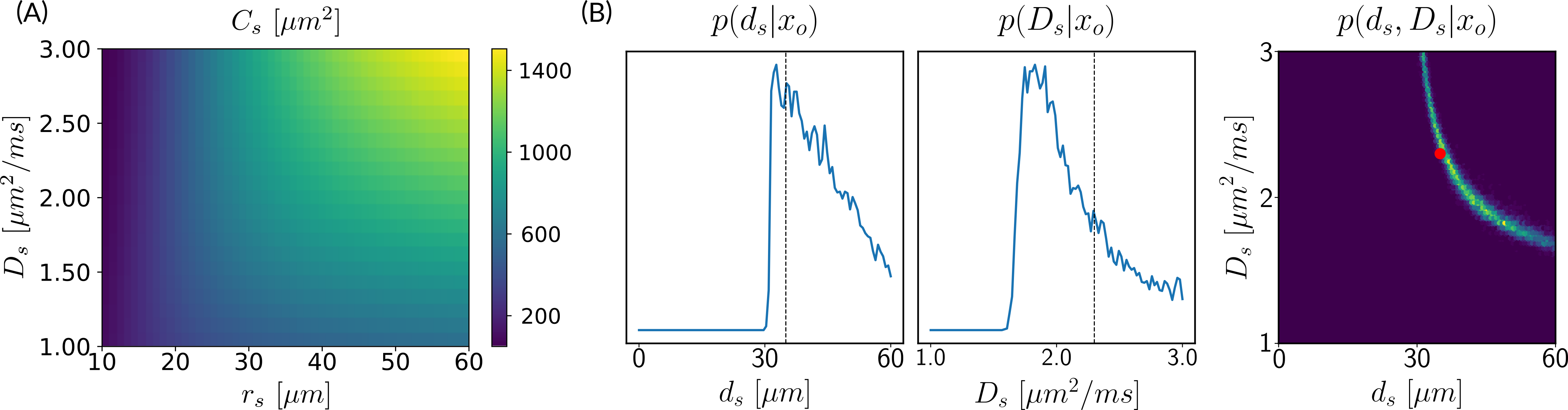}
    \caption{(A) $C_s$ dependence on soma radius $r_s$ and diffusivity $D_s$. We see that there are several values of ($d_s$, $D_s$) that yield the same $C_s$. (B) and (C) Histograms of $10^4$ samples from the marginal and joint posterior distributions of $d_s = 2r_s$ and $D_s$. The ridge in the joint distribution indicates that there are several possible values for the pair ($d_s$, $D_s$) with high probability, which are those yielding the same $C_s$. Estimating $C_s$ directly bypasses this indeterminacy.}
    \label{fig:Cs_indeterminacy}
\end{figure}

Figure \ref{fig:Cs_indeterminacy} presents the marginal posterior distributions of $r_s$ and $D_s$ as well as their joint distribution using the setup \texttt{Ideal}. To obtain these results, we have altered our LFI pipeline so to consider an extended parameter vector including $r_s$ and $D_s$. The prior distribution reflects our assumption that $r_s \in [10^{-5}, 30]$ \si{\micro\meter} and $D_s \in [1, 3]$ \si{\square\micro\meter\per\milli\second}  and we consider ground truth parameters $r_s =$ \SI{17.5}{\micro\meter} and $D_s =$ \SI{2.3}{\square\micro\meter\per\milli\second}. We note that in addition to larger marginal posterior distributions for each parameter, the joint posterior has a valley of large values for the $(r_s, D_s)$ pair, including the ground truth parameters. This result is typical of non-injective models, i.e. models for which several input parameters may yield the same output observation, and is an important asset of a probabilistic framework such as ours.

\subsubsection{Assessing the variances of estimated parameters.}
\begin{figure}
    \centering
\includegraphics[width=0.65\textheight]{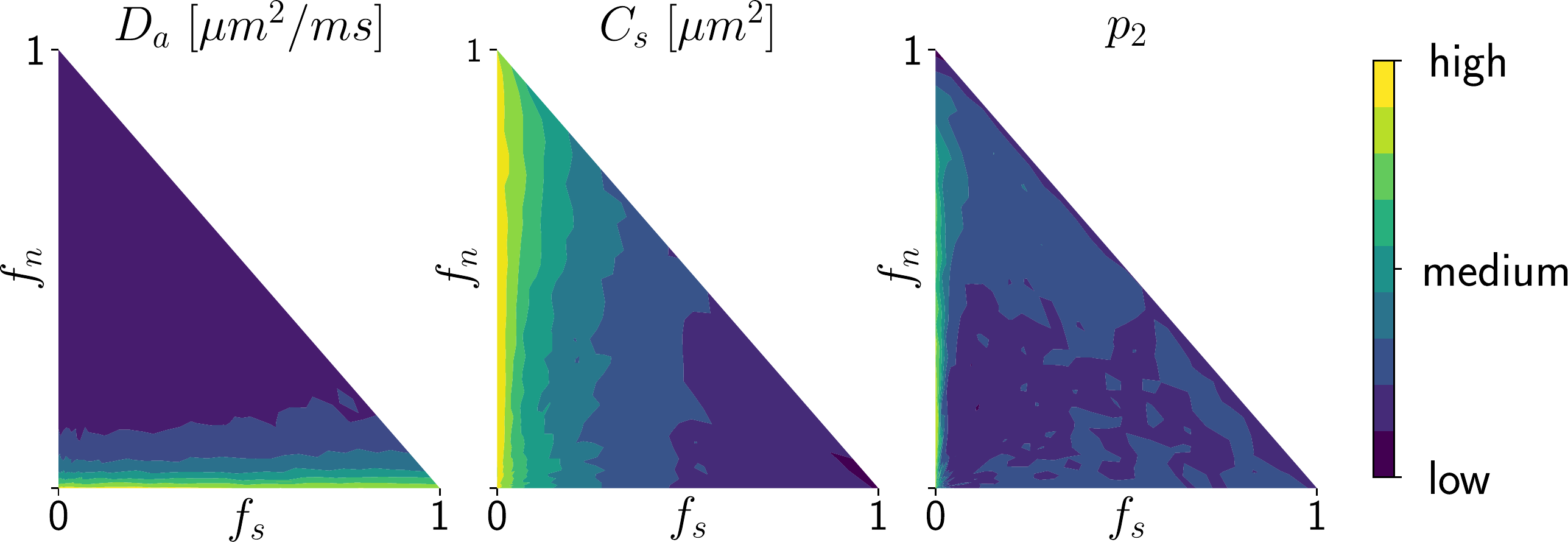}
    \caption{Logarithm of the standard deviations for the marginal posterior distribution of $D_n$, $C_s$, and $p_2$ with different choices of ground truth parameters (varying $f_s$ and $f_n$). Since the ranges of values for each plot were quite different, we labeled the colorbar in terms of \{`high', `medium', `low'\} values of standard deviations so to provide mainly qualitative information to the reader. We see that when the signal fraction of somas decreases ($f_s \to 0$) the standard deviation of the $C_s$-estimation increases; and when less neurites are present ($f_n \to 0$) the standard deviation of $p_2$ and $D_n$ increase. }
    \label{fig:dmipy_variance}
\end{figure}

Deriving the posterior distributions of the parameter vectors allows us to report the values of the most likely tissue parameters for a given observation, along with our certitude regarding our inference. Figure~\ref{fig:dmipy_variance} presents the logarithm of the standard deviation of the marginal posterior samples for different ground truth parameter choices (varying $f_s$ and $f_n$) under setup \texttt{Ideal}. These values indicate how sharp a posterior distribution is and, therefore, quantify the quality of the fit. We observe larger standard deviations in the absence (or weak presence) of soma compartments in the mixture signal, e.g., the standard deviation of $C_s$ is large when few or no somas are present ($f_s \approx 0$). This is to be expected, since the lack of contribution from the somas in the diffusion signal makes it difficult to estimate parameters related to them. 

\subsection{HCP MGH}
\label{sec:HCP_results}
Our results on simulated data show that, although we manage to invert very well the brain tissue parameter on settings for which the dMRI signal is obtained with several $b$-values, the estimates for more realistic settings are less robust and demand a more subtle analysis. Indeed, we have observed that for both setups \texttt{HCP MGH} and \texttt{HCP 1200} the soma parameters seem to be rather well estimated without too much bias, which has lead us to consider mainly these parameters in our interpretations of the results.

\begin{figure}[!htb]
    \centering
    \includegraphics[width=0.9\linewidth]{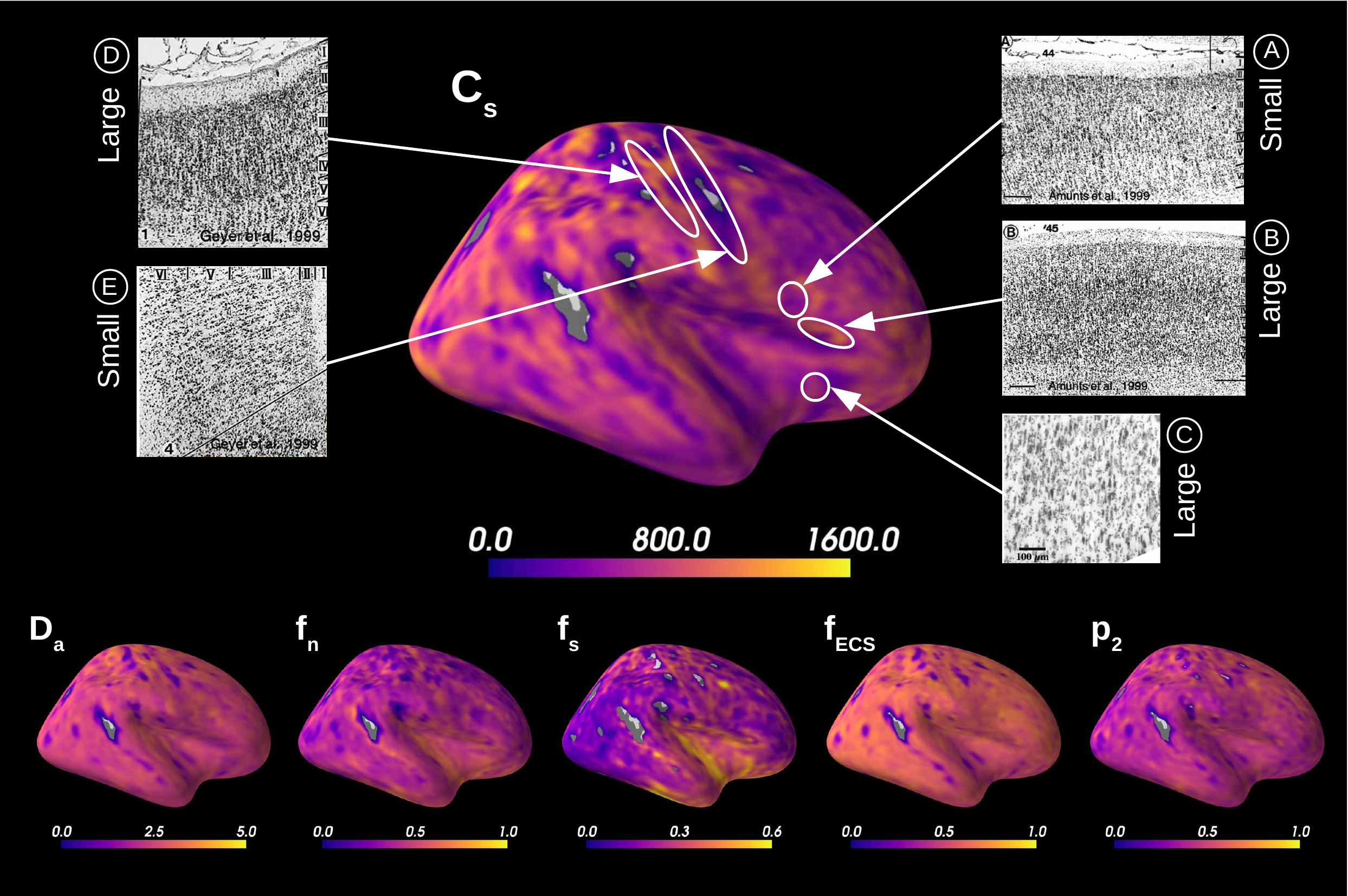}
    \caption{Microstructural measurements averaged over 31 HCP MGH subjects. We deemed stable measurements with a z-score larger than 2, where the standard deviation on the posterior estimates was estimated through our LFI fitting approach. In comparing with Nissl-stained cytoarchitectural studies we can qualitatively evaluate our parameter $C_s$: Broadmann area 44 (A) has smaller soma size in average than area 45 (B)~\citep{amuntsBrocaRegionRevisited}; large von Economo neurons predominate the superior anterior insula (C)~\citep{allmanEconomoNeuronsFrontoinsular2010}; precentral gyrus (E) shows very small somas while post-central (D) larger ones~\citep{geyerAreas3a3b1999}.
    \label{fig:HCP_results}}
\end{figure}

Figure~\ref{fig:HCP_results} presents the results on the HCP MGH dataset. 
\Review{The inference takes approximately one hour per subject using 20 CPUs, and the estimation of the seven parameters for every voxel in the grey matter is about 3 hours per subject, when computed in parallel on 20 CPUs.}
We have masked our results so to show only areas where parameters were deemed stable, i.e. when the values were larger than 2 times the LFI-obtained standard deviations of the fitted posterior, \Review{indicating that the posterior distribution is narrow and centered around its mean value}. We observe a lack of stability on small sections including the auditory cortex and the precentral gyrus fundus. Our figure assesses qualitatively the results on soma size by comparing with Nissl-stained histological studies~\citep{allmanEconomoNeuronsFrontoinsular2010,amuntsBrocaRegionRevisited,geyerAreas3a3b1999}. Our comparison shows good agreement between different cortical areas and the parameter $C_s$, which, under nearly-constant intra-soma diffusion $D_s$, is modulated by soma size. \Review{Note that we modeled brain gray matter as homogeneous per voxel. That is, we suppose only one type of neuron is present in each voxel. The most probable one is returned by the LFI approach.}

\begin{figure}[!htb]
    \centering
    \includegraphics[width=0.9\linewidth]{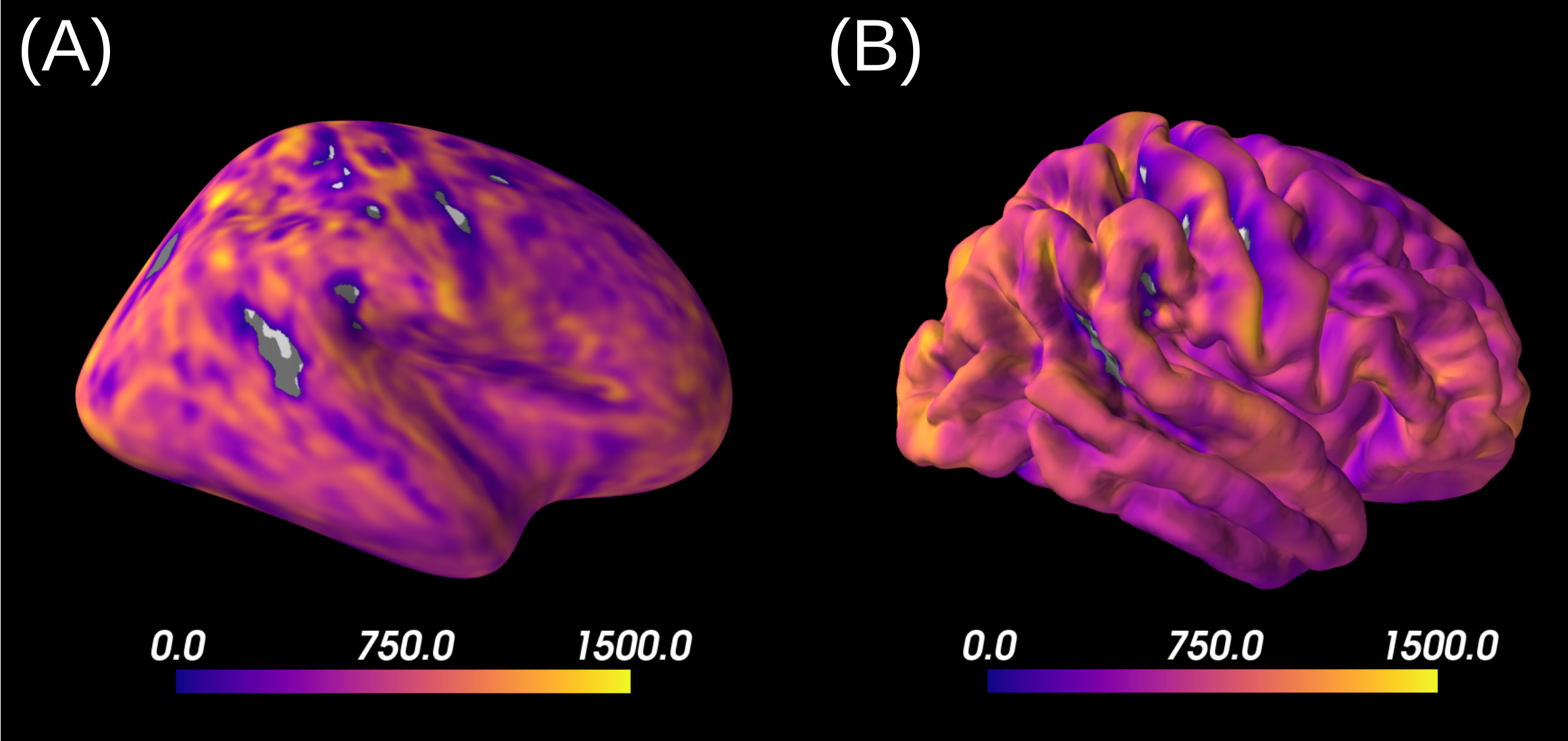}
    \caption{$C_s$ estimations averaged over 31 HCP MGH subjects, with unstable results masked, projected onto inflated (A) and pial (B) surfaces. Interestingly, low confidence areas correspond to the fundus of the sulci.
    \label{fig:HCP_pial}}
\end{figure}

{Interestingly, most regions of Figure~\ref{fig:HCP_pial} in which the parameter estimation has low confidence are located in the fundus of the sulci. Two main hypothesis could explain such behavior. Firstly, brain regions which are very curved may be more prone to mixing between tissue layers and CSF, which generates noisier signals. Thus, the estimation of summary statistics becomes \Review{more biased} and the posterior distributions tend to be wider. Another possible explanation, based on cytoarchitecture considerations, points out the fact that the fundus of sulci is where sharp changes in cellular populations happen \citep{brodmann1909vergleichende,pandya2015cerebral}. A mixing of several types of neurons in one voxel could generate multi-modal posterior distributions, and hence a region with large variance.}

\begin{figure}[!htb]
    \centering
    \includegraphics[width=0.9\linewidth]{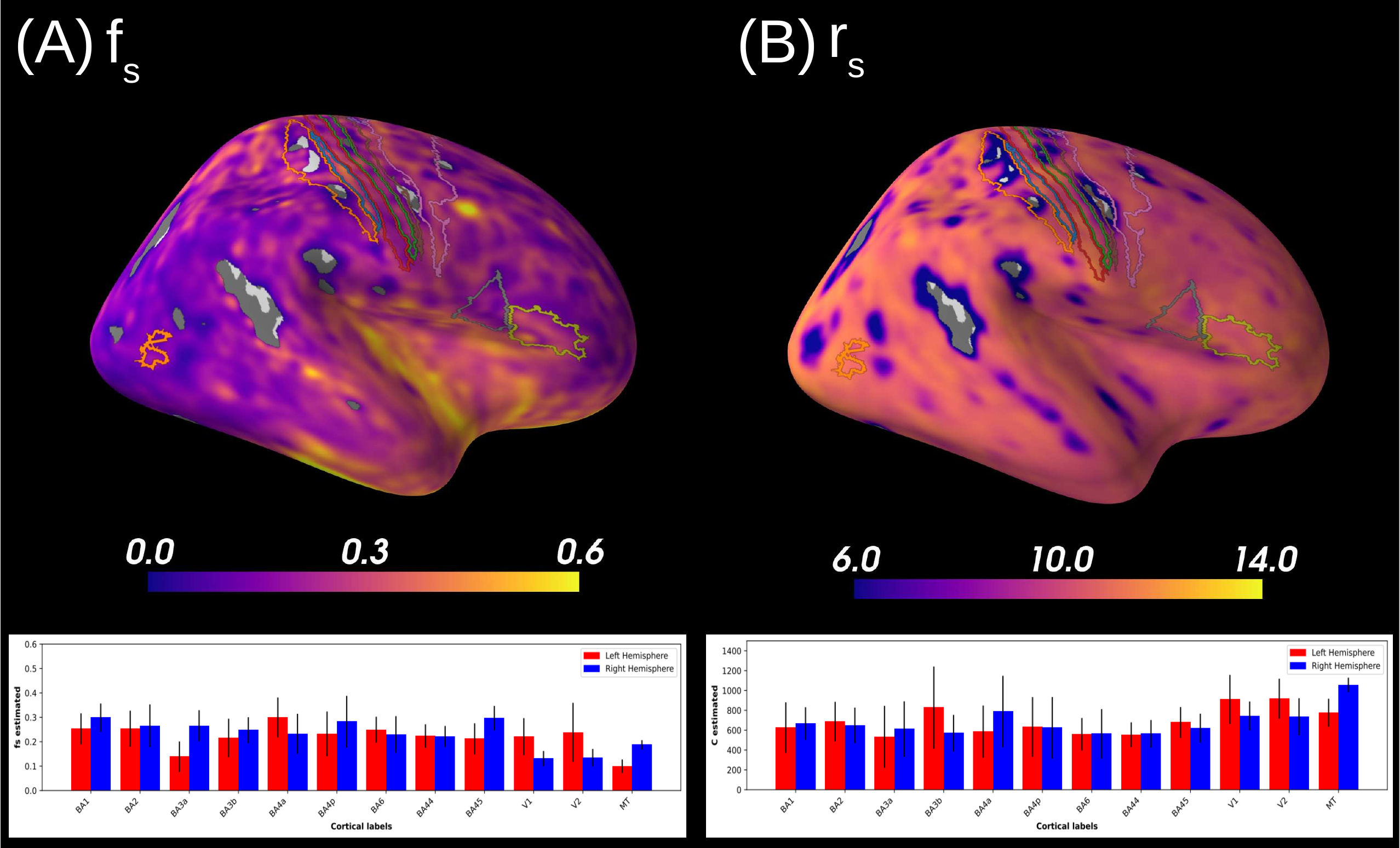}
    \caption{(Left) The average soma proportion ($f_s$) over the 31 HCP MGH subjects is projected onto an inflated cortical surface, with unstable results masked; see text for details. We also show the main Brodmann areas available on FreeSurfer. The mean values of $f_s$ over these regions (using only reliable estimations) are reported on the bar plots below. (Right) Soma radius map computed from $C_s$ with soma diffusivity fixed to $D_s = \SI{3}{\square\micro\meter\per\milli\second}$, averaged on 31 subjects. The bar plots below report the $C_s$ mean values on main Brodmann areas. We observe a good agreement between our reported values and the Brodmann areas. \label{fig:Destrieux}}
\end{figure}

Figure \ref{fig:Destrieux}A reports the soma proportion (parameter $f_s$) averaged over 31 HCP MGH subjects, masking unstable results. \Review{Mean soma signal proportion in the cortex equals $0.22$ (mean computed in trusted estimations only)}. These results are coherent with the mean volume occupancy of $~10 - 20\%$ observed in grey matter \citep{Shapson-Coe_GM_tissue}. To interpret the results at a region-based level, we have superimposed the soma proportion estimations with the Brodmann atlas. We observe a general agreement between the estimations and the atlas, and more particularly in the somatosensory and Broca's areas. A clear difference in soma proportion can be observed in the 12 regions, as presented in the barplot. 

Despite the $C_s$ parameter being useful for avoiding indeterminacies in the model inversion, its biological interpretation remains difficult. With the goal of relating our $C_s$ estimation with physiological insight, we estimated soma radius $r_s$ by fixing soma diffusivity. Similarly to the SANDI method, we fixed soma diffusivity $D_s$ to the value of the self-diffusion coefficient of free water (\SI{3}{\square\micro\meter\per\milli\second}). Note that this value could be adjusted for each voxel, and is only used here in a matter of comparison and interpretation. Using a fixed-point method, we computed the soma radius $r_s$ from the averaged $C_s$ map obtained from our posterior distribution. The results are portrayed in Figure~\ref{fig:Destrieux}B. We see that the estimated soma radius vary between $8$ and \SI{14}{\micro\meter}, which is in accordance with histology~\citep{palombo_neuron_size_2021}. Mean $C_s$ values are presented in the barplot beneath the soma radius estimations in the different Brodmann regions.

\subsection{HCP 1200}
\label{sec:HCP_1200_results}
\begin{figure}[!htb]
    \centering
    \includegraphics[width=0.4\linewidth]{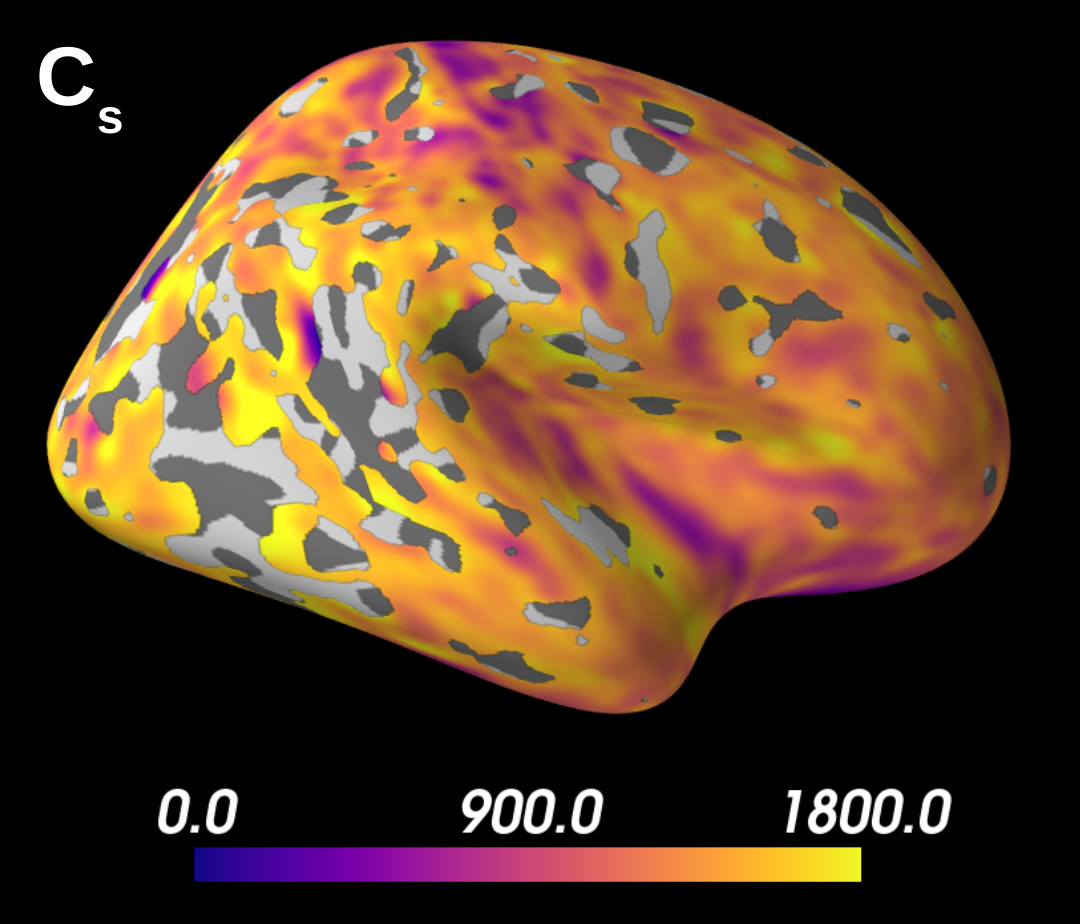}
    \caption{{Spatial distribution of the $C_s$ estimations averaged over 30 HCP 1200 subjects and projected onto an inflated cortical surface, with unstable estimations masked. Due to the scarce and rather low $b$-values ($\leq$ \SI{3}{\milli\second\per\square\micro\meter}) used in the database, the results are very unstable and, therefore, many voxels are discarded.}
    \label{fig:HCP_1200}}
\end{figure}

Figure \ref{fig:HCP_1200} shows the results obtained on a database with only three $b$-shells. We see that 55 \% of the $C_s$ estimations on brain grey matter is considered unstable and is, therefore, masked. Indeed, our $q$-bounded RTOP approximation relies on high $b$-values, where the signal is expected to have converged towards a value that depends on the radius of the soma. The larger the soma, the sooner the $q$-bounded RTOP converges. However, the largest $b$-value contained in this database equals \SI{3}{\milli\second\per\square\micro\meter}, which is not enough for the signal to have converged. \Review{Thus}, the poor quality of the summary statistics estimation leads to rather wide posterior distributions, resulting in unreliable results, as shown in simulations. Note, however, that estimations of the superior temporal gyrus for example are not masked, and both data sets seem to indicate large neurones in that region. The estimation of $D_a$ is considered as unstable for 98.7\% of the data set. This behavior was expected, given the results presented in Figure~\ref{fig:sim_base_case}, for which a large bias is observed.

\section{Discussion}

\subsection{Validation simulated data}
An important aspect of our work is the thorough validation that we have carried out on simulated data, using different acquisition setups and ways of generating diffusion signals. Part of this validation had the goal of ensuring that our method gave consistent results on simple standard cases, as confirmed by the results in Figure~\ref{fig:sim_base_case_setupA}. Additionally, we have demonstrated the benefits of using summary statistics for describing dMRI signals, attaining better parameter estimates when using them instead of directly manipulating the diffusion signals. Such results are very encouraging and should push other researchers into using these summary statistics for processing their dMRI signals.

Another relevant byproduct of our validation was observing that the quality of the parameter estimations depends heavily on the distribution of $b$-values used to acquire and simulate the dMRI signals. Indeed, if only small $b$-values are available, the summary statistics of dMRI signals are poorly estimated and the parameter estimation too. Based on this observation, we were able to apply our LFI pipeline to real datasets HCP MGH and HCP 1200 knowing in advance the limitations of our methodology; for example, we knew from which parameters we could expect good estimates (mostly soma-related ones) and which ones should not be taken into account in our analysis.


\subsection{$C_s$: A proxy to soma size}
Estimating both soma radius ($r_s$) and diffusivity ($D_s$) with diffusion MRI is a challenging task. When trying to estimate them separately, we can expect a `banana-shape' in their joint posterior distribution as shown in Figure~\ref{fig:Cs_indeterminacy}. This indicates that several values of the pair ($r_s$, $D_s$) can explain the observed signal with high probability and, therefore, one is confronted with model indeterminacy. 

The new parameter $C_s$ that we introduce in this paper is modulated both by the soma radius and its diffusivity. Thus, estimating it directly avoids problems of indeterminacy, as shown in~Figure~\ref{fig:sim_base_case_setupA}, for example. Note, however, that avoiding such indeterminacy comes with the price of losing specificity and, therefore, physiological interpretations. Fortunately, acquisitions in the narrow or wide pulse regimes allow us to better interpret estimations of $C_s$, as it only depends on $r_s$ or $D_s$ (see Section~\ref{sec:3-compartment-model}). 

\subsection{Can I apply this approach to my data?}

One of the main benefits of a probabilistic framework is that it can be applied to any data set or acquisition setup without too much hesitation, since the estimates always come with a ``quality certificate'' described by the credible regions derived from the posterior distribution. We have benefited from this feature when creating all figures related to databases HCP MGH and HCP 1200, since they allow us to mask regions for which the variance of the parameter estimation is too high. We can also use it as a proxy to identify regions for which the three-compartment model is adequate or not, or assessing whether the distribution of $b$-values used to acquire the observed diffusion signal is sufficiently informative. \Review{Note, however, that it would not be realistic to expect that our method should give acceptable results on every database to which it is applied. Indeed, the distribution of $b$-values used to acquire the data under study is a key predictor of whether the estimated posterior distribution will be useful for inverting the three-compartment model or not. For instance, if the $b$-values are too small, then the RTOP summary statistics will be poorly estimated in most voxels, leading to unreliable estimations; only regions with larger somas will be correctly estimated. Similarly, if the acquisition uses only $b$-values much greater than zero, the Spiked LEMONADE moments will be biased, which also leads to neurite estimations with large variability. These observations were useful when analyzing the \texttt{HCP 1200} database, since the distribution of $b$-values are concentrated at low values and, therefore, our estimations are prone to bias and have large variances.}



\subsection{Limitations and perspectives}
There are many extensions that we could envision for the method that we propose.


\Review{The proposed approach has been designed and applied to brain grey matter, but one could also want to apply it to brain white matter (see Supplementary Material). Results on the HCP MGH dataset indicate that the axons distribution is more anisotropic in white matter than grey matter ($p_{2,WM} > p_{2, GM}$), with values coherent to the ones obtained with the LEMONADE framework \citep{novikov_rotationally-invariant_2018}. Somas are also less present in white matter than grey matter, as indicated by a lower signal fraction $f_s$, which is expected from histology \citep{Shapson-Coe_GM_tissue}. However, the ECS model used in the three-compartment model defines its diffusivity as isotropic. While this assumption seems to hold in brain gray matter, a tensor representation is usually preferred for the ECS diffusivity, as it is the case in the SM \citep{novikov_quantifying_2018}. The application of a similar LFI pipeline based on a model designed for brain white matter (such as the SM) could help improve the estimations of tissue parameters and better interpret its output thanks to the posterior distributions. As an example, this pipeline could be a new way to solve the LEMONADE system of equations proposed by \citet{novikov_rotationally-invariant_2018} within a probabilistic framework.}

\Review{The three-compartment model that we use approximates well the intra-cellular signal in brain grey matter tissue by adding a sphere compartment to account for soma presence \citep{PALOMBO_SANDI_2020}. However, this is a rather simplified model and it could be improved; for instance, the geometry of ECS is very restricted and tortuous, and diffusion signals have been proven to deviate from a mono-exponential behavior \citep{Vincent_sucrose_mouse_2021}. A first improvement could be, as mentioned before, to deviate from modelling the ECS diffusion with a simple isotropic Gaussian and consider more complex geometric representations. Another improvement would be to consider exchanges between the neurites and the ECS \citep{jelescu_NEXI_2021,olesen2021diffusion}.}

\Review{In this method, we require to have a signal with large $q$-values to extract summary statistics using RTOP. This constraint limits the number of possible applications, as many dMRI datasets do not contain such high $q$-values. A solution proposed by \citet{golkov2016qspaceDL} is to rely on deep learning to learn a mapping between a limited number of DWIs and some microstructure scalars, assuming that all the necessary information is contained in the low-dimensional acquisition. Following this $q$-space deep learning method, an improvement of the proposed method could be to learn the summary statistics from a dataset with less $q$-shells using deep learning, and then apply our likelihood-free inference method to solve the inverse problem. 
} 

Finally, we could leverage the modularity of the probabilistic framework and apply our methodology to other kinds of models in medical imaging. This could be done by simply re-defining the forward model and the summary statistics defined in Section~\ref{sec:lfi-setup}. Such flexibility confirms the generality of our approach and the potential impact that it might have in the medical imaging field.  

\section{Conclusion}
Quantifying grey matter tissue composition is challenging. In this work, we have presented a methodology to estimate the parameters of a model that best fit an observed data point, and also their full posterior distribution. This rich description provides many useful tools, such as assessing the quality of the parameter estimation or characterizing regions in the parameter space where it is harder to invert the model. The inclusion of such ``quality certificate'' accompanying our parameter estimation is very useful in practice and allows one to apply the pipeline on any kind of database and know to which degree one can trust the results. Moreover, our proposal alleviates limitations from current methods in the literature by not requiring physiologically unrealistic constraints on the parameters and avoiding indeterminacies when estimating them.

To conclude, we believe that our approach based on Bayesian inference with modern tools from neural networks is a promising one that can easily be applied to other applications in the medical imaging field: one only needs to define a sufficiently rich model describing a certain phenomenon of interest and the LFI pipeline will manage to invert it and provide a related posterior distribution. We expect, therefore, that other researchers will find this contribution valuable for their own applications and see such probabilistic approach more often used in the literature.


\acks{
Experiments on diffusion MRI data were made possible thanks to DiPy~\citep{dipy2014}, as well as the scientific Python ecosystem: Python~\citep{Python36}, Matplotlib~\citep{hunter2007matplotlib}, Numpy~\citep{harris2020array}, Scipy~\citep{2020SciPy-NMeth}, Mayavi~\citep{ramachandran2011mayavi}, SBI~\citep{tejero-cantero2020sbi} and PyTorch~\citep{Paszke2019pyTorch}.

This work was supported by grants ERC-StG NeuroLang ID:757672 and ANR-NSF NeuroRef to DW and the ANR BrAIN (ANR-20-CHIA-0016) grant to AG.}

%
\ethics{The work follows appropriate ethical standards in conducting research and writing the manuscript, following all applicable laws and regulations regarding treatment of animals or human subjects.}

\coi{We declare we don't have conflicts of interest.}

\bibliography{mylib}

\newpage
\appendix 
\section*{Appendix A.}
	\Review{In this appendix we present additional experimental results of brain white matter microstructure estimations using the proposed method.}
	
\begin{figure}[!htb]
        \centering
        \includegraphics[width=\linewidth]{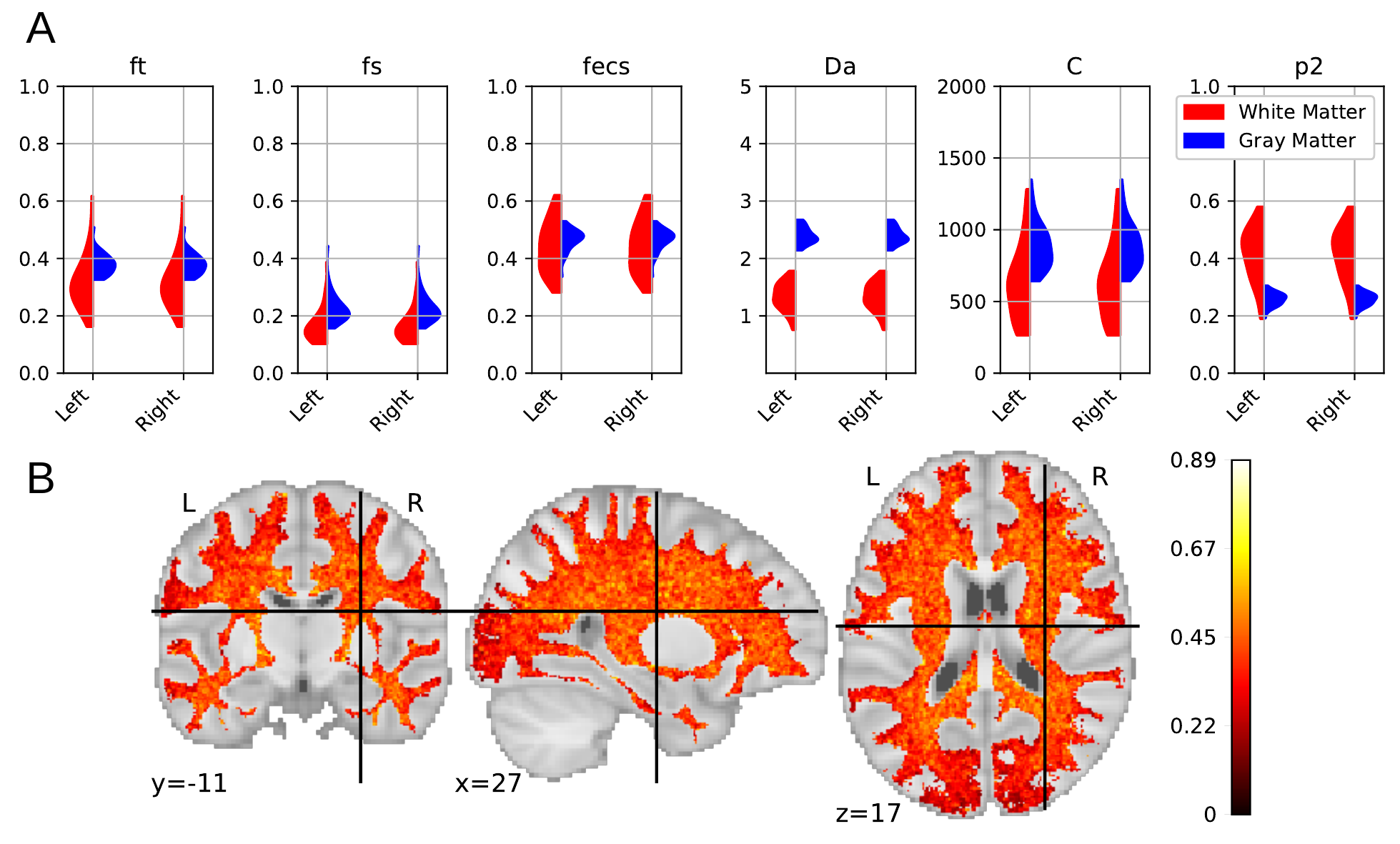}
        \caption{\Review{Mean estimation over HCP MGH subjects of brain white matter microstructure parameters using LFI methods on a three-compartment model. A. Higher $p_2$ values are observed in brain white matter compared to gray matter, indicating a more anisotropic axon distribution. Soma proportion is also reduced compared to gray matter estimations, along with smaller soma size. These estimations, although encouraging, should be taken with caution, because the ECS model used here is not suited for white matter tissue. B. $p_2$ estimations in brain white matter, 0 indicating an isotropic distribution of axons, and 1 an anisotropic distribution (i.e. perfectly aligned fibers).}
        \label{fig:WM}}
    \end{figure}

	\Review{Figure \ref{fig:WM} A presents a comparison of the mean estimations of white matter and gray matter parameters among all HCP MGH subjects, keeping only trusted voxels. We can observe a lower proportion of somas in white matter ($f_{s,\text{WM}} < f_{s,\text{GM}}$) along with smaller soma size compared with grey matter ($C_{s,\text{WM}} < C_{s,\text{GM}}$). This weak presence of small somas in white matter is in accordance with histology. These results also indicate a more anisotropic distribution of axons is white matter than neurites in grey matter ($p_{2,\text{WM}} > p_{2, \text{GM}}$). Obtained $p_2$ values are coherent to the ones obtained with the LEMONADE framework \citep{novikov_rotationally-invariant_2018}. Figure \ref{fig:WM} B presents the mean $p_2$ values in white matter averaged over all subjects, keeping only trusted estimations. The center of brain white matter appears more anisotropic than at the frontier with grey matter.}
	
	\Review{In this paper we approximate the ECS with an isotropic diffusion. Models specific to brain white matter, such as the SM, usually represent ECS diffusivity as a non-isotropic tensor \citep{novikov_quantifying_2018}. Results obtained from this model should then be taken with caution, because this model does not reflect white matter tissue properly.}
	
\end{document}